\documentclass[12pt]{article}

\usepackage{amssymb}
\usepackage{amsmath}
\usepackage{amscd}
\usepackage{latexsym}
\usepackage{graphicx}

\usepackage{cite}

\newcommand{\be}{\begin{equation}}
\newcommand{\ee}{\end{equation}}
\newcommand{\Dlt}{\Delta}
\newcommand{\dlt}{\delta}

\newcommand{\br}{{\bf r}}
\newcommand{\ba}{{\bf a}}
\newcommand{\bfe}{{\bf e}}
\newcommand{\bn}{{\bf n}}
\newcommand{\bS}{{\bf S}}

\newcommand{\bB}{{\bf B}}
\newcommand{\bt}{\beta}

\newcommand{\al}{\alpha}

\newcommand{\gm}{\gamma}
\newcommand{\om}{\omega}
\newcommand{\Om}{\Omega}

\newcommand{\dgr}{\dagger}

\newcommand{\rgl}{\rangle}
\newcommand{\lgl}{\langle}

\begin{document}

\begin{center}
 
{\Large{\bf Regulating spin dynamics of dipolar and spinor atoms} \\ [5mm]

V.I. Yukalov$^{1,2}$ and E.P. Yukalova$^{3}$}  \\ [3mm]

{\it
$^1$Bogolubov Laboratory of Theoretical Physics, \\
Joint Institute for Nuclear Research, Dubna 141980, Russia \\ [2mm]

$^2$Instituto de Fisica de S\~ao Carlos, Universidade de S\~ao Paulo, \\
CP 369, S\~ao Carlos 13560-970, S\~ao Paulo, Brazil \\ [2mm]

$^3$Laboratory of Information Technologies, \\
Joint Institute for Nuclear Research, Dubna 141980, Russia } \\ [3mm]

{\bf E-mails}: {\it yukalov@theor.jinr.ru}, ~~ {\it yukalova@theor.jinr.ru} \\

\end{center}

\vskip 1cm

\begin{abstract}

Systems of atoms or molecules can possess nonzero total spins that can be employed
in spintronics, for instance for creating memory devices, which requires to be able
to efficiently regulate spin directions. This article presents some methods of 
regulating spin dynamics, developed by the authors. These methods can be applied
for different physical systems, described by different Hamiltonians including the
interactions of spins with magnetic fields. The principal feature of the methods
is the use of the feedback fields formed by moving spins.   

\end{abstract}

\vspace{1cm}

\section{Introduction}

One of the most important problems in spintronics is the ability of fast regulation
of spin dynamics. This ability is indispensable for the creation of different 
devices, such as memory storages, using the interaction of effective system spins 
with magnetic fields. The functioning of memory devices usually confronts the 
necessity of overcoming two contradictory requirements. From one side, for keeping
the memory for long time, it is necessary to be able to freeze the spin direction,
while from the other side, it is required, when necessary, to quickly change this
direction. 

In this communication, we describe some methods, developed by the authors, allowing 
for fast regulation of spin motion. These methods can be applied to different systems
enjoying magnetic moments. In order to achieve coherent spin dynamics, it is customary
to employ nanosize objects, such as magnetic nanomolecules \cite{Barbara_1,Caneschi_2,
Yukalov_3,Yukalov_4,Yukalov_5,Yukalov_6,Yukalov_7,Yukalov_8,Henner_9,Henner_10,
Friedman_11,Yukalov_12,Miller_13,Craig_14,Yukalov_15,Liddle_16,Rana_17,Yukalov_18,
Yukalov_19}, magnetic nanoclusters \cite{Yukalov_7,Yukalov_12,Yukalov_15,Yukalov_18,
Yukalov_19,Kodama_20,Hadjipanayis_21,Wernsdorfer_22,Yukalov_23,Yukalov_24,Kharebov_25,
Kudr_26,Yukalov_27}, magnetic graphene, where magnetism is induced by magnetic defects
\cite{Yaziev_28,Terrones_29,Bekyarova_30,Enoki_31,Yukalov_32,Yukalov_33}, trapped atoms
and molecules interacting through dipolar or spinor forces \cite{Griesmaier_34,
Baranov_35,Baranov_36,Stamper_37,Gadway_38,Yukalov_39,Yukalov_40}, magnetic quantum 
dots \cite{Schwartz_41,Birman_42,Mahajan_43,Tufani_44}, polarized nanomolecules 
\cite{Yukalov_45,Yukalov_46}, and magnetic hybrid materials \cite{Odenbach_47}.

The method we have suggested is based on the use of the combination of a magnetic 
material and an electric circuit creating a magnetic feedback field. This method can 
be applied to various magnetic materials. Of course, different materials are described
by different Hamiltonians, which requires to employ some special tricks for being able
to effectively regulate their spin motion, however the basic ideas remain the same.

\section{Magnetic nanomolecules and nanoclusters}

The main ideas employed for regulating spin dynamics can be most easily explained by
the example of spin dynamics in magnetic nanomolecules and nanoclusters. 

Magnetic nanomolecules have the sizes of nanometers. A molecule potential topography 
can be described by a double-well potential, where the spin possesses two easy directions
that can be called "spin up" and "spin down". Below the blocking temperature 
$T_B \sim 1-10$ K, the spin becomes frozen in one of the directions due to the magnetic
anisotropy field $E_A \sim 10-100$ K. The total spin of the molecule ground state can
vary between $1/2$ and $27/2$. Thus the widely studied molecules of Mn$_{12}$ or Fe$_8$
have the spin $S = 10$.

Magnetic clusters are characterized by the similar properties, although they are much 
larger than molecules, being of the radius $R_{coh} \sim 10-100$ nm and containing
about $100-10^5$ magnetic particles. Below the blocking temperature $T_B \sim 10-100$ K,
the cluster spin is frozen in one of the easy directions. Magnetic nanoclusters have to
be of nanosizes mentioned above in order that the spins of particles forming the clusters
be coherent with each other and the specimen would not separate into several domains.
Magnetic particles of the sizes $R > R_{coh}$ become separated into several magnetic 
domains, so that the total spin becomes close to zero. There exists a very large variety
of magnetic clusters composed of pure metals, such as Fe, Ni, and Co, and of various 
oxides and alloys. 

The dilemma, one confronts when creating memory devices, is as follows. From one side,
to keep the fixed memory intact for long time, one needs to possess a rather strong 
magnetic anisotropy, while from the other side, to be able to quickly change the spin 
direction, in order to erase the memory or to correct the stored content, one has to   
be able to quickly regulate spin dynamics.

Magnetic nanomolecules and nanoclusters are described by the similar Hamiltonians,
such as
\be
\label{1}
\hat H = - \mu_S \bB \cdot \bS - D S_z^2 + E ( S_x^2 - S_y^2 ) \;  ,
\ee
where $\mu_S$ is magnetic moment, $\bf B$ is the total magnetic field acting on the
specimen, $\bS$ is the spin operator, $D$ and $E$ are anisotropy parameters. The total
magnetic field is the superposition 
\be
\label{2}
\bB = ( B_0 + \Dlt B ) \bfe_z + H \bfe_x + B_1 \bfe_y
\ee
of a constant magnetic field $B_0$, regulated field $\Delta B$, anisotropy field $B_1$,
and a feedback field $H$. 

The overall setup is organized according to the scheme of Fig. 1, where the studied 
magnetic sample is inserted into a magnetic coil of an electric circuit playing the role 
of a resonator. The spin is in a metastable state, being frozen due to the low temperature
and strong magnetic anisotropy. The moving spin of the sample creates electric current 
in the circuit that forms the feedback field acting on the sample. The feedback field 
satisfies the equation
\be
\label{3}
\frac{dH}{dt} + 2\gm H + \om^2 \int_0^t H(t') \; dt'= - 4\pi \eta_{res} \;
\frac{dm_x}{dt} \;   ,
\ee
in which $\gamma$ is the circuit attenuation, $\omega$ is the circuit natural frequency,
$\eta_{res} = V/V_{res}$ is the resonator filling factor, and the electromotive force is 
created by the moving average spin
\be
\label{4}
m_x = \frac{\mu_S}{V} \; \lgl \; S_x \; \rgl \; .
\ee
   
The equations of motion are written for the spin averages
\be
\label{5}
x \equiv \frac{ \lgl \; S_x \; \rgl }{S} \; , \qquad
y \equiv \frac{ \lgl \; S_y \; \rgl }{S} \; , \qquad
z \equiv \frac{ \lgl \; S_z \; \rgl }{S} \;   .
\ee
Pair spin correlations are decoupled employing the generalized mean-field approximation
\cite{Yukalov_6}
\be
\label{6}
\lgl \; S_\al S_\bt + S_\bt S_\al \; \rgl  = \left( 2 - \; \frac{1}{S} \right)
\lgl \; S_\al \; \rgl \lgl \; S_\bt \; \rgl \;  ,
\ee
where $\alpha \neq \beta$ and which is exact for $S = 1/2$ and asymptotically exact for 
$S \gg 1$. 

Let us introduce the notations for the Zeeman frequency
\be
\label{7}
 \om_0 \equiv -\; \frac{\mu_S}{\hbar} \; B_0  
\ee
and the anisotropy frequencies
\be
\label{8}
 \om_D \equiv ( 2S - 1 ) \; \frac{D}{\hbar} \; , \qquad 
\om_E \equiv ( 2S - 1 ) \; \frac{E}{\hbar} \; , \qquad 
\om_1 \equiv  -  \; \frac{\mu_S}{\hbar} \; B_1 \;  .
\ee
The dimensionless anisotropy parameter is
\be
\label{9}
 A \equiv \frac{\om_D + \om_E}{\om_0} \;  .
\ee
The dimensionless regulated field is defined as
\be
\label{10}
b \equiv -\; \frac{\mu_S\Dlt B}{\hbar\om_0} \;   .
\ee
The coupling between the magnetic sample and the resonator is characterized by the 
coupling rate
\be
\label{11}
 \gm_0 \equiv \pi \eta_{res} \; \frac{\mu_S^2 S}{\hbar V} \;   .
\ee
Also, introduce the notation for the dimensionless feedback field
\be
\label{12}
h \equiv -\; \frac{\mu_S H}{\hbar\gm_0 } \;  .
\ee
Then the equations of spin motion read as 
$$
\frac{dx}{dt} = - \om_S y + \om_1 z \; , \qquad 
\frac{dy}{dt} = \om_S x - \gm_0 h z \; ,
$$
\be
\label{13}
 \frac{dz}{dt} = 2 \om_E x y - \om_1 x + \gm_0 h y  \;  ,
\ee
where
\be         
\label{14}
\om_S \equiv \om_0 ( 1  + b - Az )
\ee
plays the role of an effective rotation frequency.

In the dimensionless notation, the feedback-field equation becomes
\be
\label{15}
 \frac{dh}{dt} + 2\gm h + \om^2 \int_0^t h(t') \; dt'=
4 \; \frac{dx}{dt} \;  .
\ee

Efficient interaction between the electric circuit and the magnetic sample occurs 
when there happens a resonance between the effective Zeeman frequency $\om_S$ of 
the sample spins and the resonator natural frequency $\omega$. However, because 
of the presence of the anisotropy term, such a resonance cannot occur, since the 
detuning
\be
\label{16}
\frac{\om_S -\om}{\om_0}  = \frac{\Dlt\om}{\om_0} + b - A z \qquad
(\Dlt \om \equiv \om_0 - \om )
\ee
can be very large.

One of the possibilities for overcoming the above problem is to resort to the method 
of {\it triggering resonance} \cite{Yukalov_18}. Assume that we need to overturn the 
spin at the moment of time $\tau$. Then, tuning $\omega$ to $\omega_0$, the regulated 
part of the magnetic field $b(t)$ is kept zero before the time $\tau$, and is switched 
on at the time $\tau$, so that
\begin{eqnarray}
\label{17}
b(t)  = \left\{ \begin{array}{ll}
0 \; ,     ~ & ~ t < \tau \\
A z_0 \; , ~ & ~ t \geq \tau
\end{array} 
\right. \; .
\end{eqnarray}

At this initial moment of time $\tau$, when the spin polarization is $z_0$, the effective 
detuning becomes small, 
\be
\label{18}
\frac{\om_S -\om}{\om_0} = b(\tau) - Az_0 = 0 \qquad ( t = \tau) \;   ,
\ee
which triggers the fast motion of spin reversal. Although the overall reversal is rather 
fast, but there appear tails, as is seen in Fig. 2.

To preserve the resonance condition for the whole time of spin reversal, the method of
{\it dynamic resonance tuning} is suggested \cite{Yukalov_19}. Then the regulated field
$b = b(t)$ is varied so that to compensate the temporal variation of the anisotropy term
$Az =  Az(t)$. That is, the regulated field is switched on so that 
\begin{eqnarray}
\label{19}
b(t)  = \left\{ \begin{array}{ll}
0 \; ,            ~ & ~ t < \tau \\
A z_{reg}(t) \; , ~ & ~ t \geq \tau
\end{array} 
\right. \; ,
\end{eqnarray}
which makes the effective detuning close to zero,
\be
\label{20}
 \frac{\om_S -\om}{\om_0} = A [ \; z_{reg}(t) - z(t) \; ] \ll 1 \;  ,   
\ee
when $\omega = \omega_0$. This method provides ideal conditions for keeping the spin 
frozen before the required time $\tau$ and then, at time $\tau$, reversing the spin, 
as shown in Fig. 3.

\section{Assemblies of nanomolecules or nanoclusters}  

It is possible to consider assemblies of nanomolecules or nanoclusters, when the system
Hamiltonian  has the form
\be
\label{21}
\hat H = - \mu_S \sum_i \bB \cdot \bS_i + \hat H_A + \hat H_D \;  ,
\ee
where the first term is the Zeeman energy, the second term is the anisotropy energy
\be
\label{22}
 \hat H_A = - \sum_j \left\{ D(S_j^z)^2 - E \; \left[ \; ( S_j^x )^2 -
( S_j^y)^2 \; \right] \right\} \;  ,
\ee
and, in addition, it is necessary to take account of the dipolar interactions through
the dipolar tensor $D_{ij}^{\alpha \beta}$,
\be
\label{23}
 \hat H_D = \frac{1}{2} \sum_{i\neq j} \;
\sum_{\al\bt} D_{ij}^{\al\bt} S_i^\al S_j^\bt \; .
\ee
The total magnetic field can be taken as
\be
\label{24}
\bB = ( B_0 + \Dlt B ) \bfe_z + H \bfe_x \;   .
\ee
The anisotropy parameter $E$ usually is much smaller than $D$, because of which one 
takes the anisotropy term as
\be
\label{25}
 \hat H_A = - \sum_j D ( S_j^z )^2 \;  .
\ee
     
The dynamics of the summary spin is similar to that of the spins of single nanosamples
\cite{Yukalov_5,Yukalov_7,Yukalov_8,Henner_9,Henner_10,Yukalov_12,Yukalov_15,Yukalov_23,
Yukalov_24,Kharebov_25,Yukalov_27}. The role of the dipolar interactions is two-fold.
First, the dipolar spin interactions trigger the initial spin motion by creating 
spin waves, and second, they result in the transverse spin attenuation leading to 
the dephasing of spin dynamics. The latter, however, does not strongly influence the 
coherent spin rotation, if the dephasing time is much longer than the reversal time. 
For nanomolecules and nanoclusters, the reversal time is of the order
\be
\label{26}
t_{rev} \approx \frac{\gm}{\gm_0\om_0 z_0} \sim 10^{-11}\; {\rm s} \;   .
\ee

\section{Other magnetic nanomaterials}

Except nanomolecules and nanoclusters, there are several other types of magnetic 
nanomaterials that can be employed for spintronic devices. Coherent spin motion 
in these materials can be regulated similarly to the case of nanomolecules and 
nanoclusters, considered above. 

Emergence of defect-induced magnetism in graphene materials has been studied by 
Yazyev \cite{Yaziev_28} and by Enoki and Ando \cite{Enoki_31}. Defects in graphene 
interact with each other through exchange interactions. The spin Hamiltonian consists 
of two terms, the Zeeman energy and the Heisenberg Hamiltonian,
\be
\label{27}
 \hat H = -\mu_S \sum_i \bB \cdot \bS_i - \; \frac{1}{2}
\sum_{i\neq j} J_{ij} \left( S_i^x S_j^x + S_i^y S_j^y +
\al S_i^z S_j^z \right) \; .
\ee
The total magnetic field is the sum as in (\ref{24}), including an external magnetic 
field and a feedback magnetic field. Peculiarities in the case of exchange interactions 
have been considered in detail in Ref. \cite{Yukalov_48}. Dynamics of defect spins in 
graphene are considered in Refs. \cite{Yukalov_32,Yukalov_33}. Similar dynamics is 
exhibited by spins of quantum dots and some magnetic hybrid materials. 

One more class of magnetic materials is presented by polarized nanomolecules that do 
not possess spins in their ground state, but nuclear spins inside them can be polarized 
by means of dynamic nuclear polarization \cite{Yukalov_49,Yukalov_50,Yukalov_51}. This 
kind of molecules, for instance, are propanediol, butanol, and ammonia. Being spin 
polarized, they can keep this polarization for days and months. The Hamiltonian is
\be
\label{28}
 \hat H = -\mu_0 \sum_i \bB \cdot \bS_i + \hat H_D \;  ,
\ee
in which the term due to dipolar interactions is
\be
\label{29}
 \hat H_D = \frac{1}{2} \sum_{i\neq j} \frac{\mu_0^2}{r_{ij}^3} \;
\left[ \; \bS_i \cdot \bS_j - 3 ( \bS \cdot \bn_{ij} ) 
  ( \bS_j \cdot \bn_{ij}) \;\right] \; ,
\ee
where
$$
 \bn_{ij} \equiv \frac{\br_{ij}}{r_{ij}} \; , \qquad 
\br_{ij} \equiv \br_i - \br_j \; , \qquad r_{ij} \equiv |\; \br_{ij} \; | \; .
$$
The dipolar Hamiltonian (\ref{29}) can be represented in the form (\ref{23}), with 
the dipolar tensor
\be
\label{30}
D_{ij}^{\al\bt} = \frac{\mu_0^2}{r_{ij}^3} \; \left( \dlt_{\al\bt} -
3 n_{ij}^\al n_{ij}^\bt \right) \;   .
\ee

The binary mixture of nuclear and electron spins makes the effective spin-resonator 
coupling larger, thus diminishing the reversal time \cite{Yukalov_6,Yukalov_52,Yukalov_53}.

\section{Trapped spinor atoms}

Many atoms and molecules possessing dipolar or angular momenta can be confined in traps 
forming trapped clouds \cite{Griesmaier_34,Baranov_35,Baranov_36,Stamper_37,Gadway_38,
Yukalov_39,Yukalov_40}. The derivation of an effective spin Hamiltonian for these 
objects goes in the following steps \cite{Yukalov_39}.

Let us assume that an atom has angular momentum
\be
\label{31}
{\bf F} = [\; {\bf F}_{mn} \; ] = 
\sum_\al [\; F_{mn}^\al \; ] \bfe_\al \;  ,
\ee
where $\alpha = x,y,z$, the quantum magnetic number takes the values 
$m = -F,-F+1,-F+2,\ldots,F-2,F-1,F$. The field operators of atoms are the columns in 
the space of the magnetic number,
$$
\psi(\br,t) = [\; \psi_m(\br,t) \; ] \;   .
$$

The Hamiltonian of a system composed of such atoms consists of several terms:
\be
\label{32}
\hat H = \hat H_0 + \hat H_{LZ} + \hat H_{QZ} + \hat H_F + \hat H_D \; ,
\ee
where
\be
\label{33}
\hat H_0 = \int \psi^\dgr(\br) \left[\; - \frac{\nabla^2}{2m} + 
U(\br) \; \right] \; \psi(\br) \; d\br
\ee
is a single-atom term without a magnetic field, the second part is a linear Zeeman 
term, the third part is a quadratic Zeeman term, the fourth part represents local 
atomic interactions, and the last part describes dipolar interactions. For the 
compactness of the formulas, here and in what follows, we set the Planck constant 
$\hbar$ to one.

The linear Zeeman term is the standard expression
\be
\label{34}
 \hat H_{LZ} = - \mu_F 
\int \psi^\dgr(\br)\; \bB \cdot {\bf F} \; \psi(\br) \; d\br \;   ,
\ee
where $\mu_F$ is the atomic magnetic moment. 

The quadratic Zeeman term consists of the sum
\be
\label{35}
 \hat H_{QZ} = Q_Z
\int \psi^\dgr(\br)\; (\bB \cdot {\bf F})^2 \; \psi(\br) \; d\br +
q_Z  \int \psi^\dgr(\br)\; ( F_z)^2 \; \psi(\br) \; d\br \;  \;    ,
\ee
in which the first term in the write-had site describes static-current quadratic 
Zeeman effect and the second term, alternating-current quadratic Zeeman effect. 

The nonresonant static-current quadratic Zeeman effect arises in atoms possessing 
hyperfine structure, hence a nonzero nuclear spin 
\cite{Jenkins_54,Schiff_55,Killingbeck_56,Coffey_57,Woodgate_58,Demtroder_59}. The 
static-current quadratic Zeeman effect parameter reads as
\be
\label{36}
 Q_Z = \mp \; \frac{\mu_F^2}{\Dlt W(1+2I)^2} \;  ,
\ee
where $\Delta W$ is a hyperfine energy splitting and $I$ is nuclear spin. The sign minus 
or plus in the static-current Zeeman parameter $Q_Z$ is defined by the relative alignment 
of the nuclear and the total electron spin projections of the atom: minus for parallel
projections, while plus for antiparallel projections.

The quasi-resonant alternating-current quadratic Zeeman effect is due to the 
alternating-current Stark shift caused either by applying a linearly polarized microwave 
driving field inducing hyperfine transitions \cite{Gerbier_60,Leslie_61,Bookjans_62} 
or by applying off-resonance linearly polarized light inducing transitions between 
internal spin states \cite{Cohen_63,Santos_64,Jensen_65,Paz_66}. The alternating-current
quadratic Zeeman effect parameter is
\be
\label{37}
 q_Z = - \; \frac{\Om_R^2}{4\Dlt} \;  ,
\ee 
where $\Omega_R$ is the Rabi frequency of the driving alternating field and $\Delta$ 
is the detuning from an internal, spin or hyperfine, transition.  

Local atomic interactions are presented by the Hamiltonian
\be
\label{38}
 \hat H_F = \frac{1}{2} \sum_{klmn} \int \psi_k^\dgr(\br) \;  \psi_l^\dgr(\br) 
\Phi_{klmn} \;  \psi_m(\br)  \psi_n(\br) \; d\br d\br' \;   ,
\ee
in which $\Phi_{klmn}$ is a matrix element of the interaction potential
\be
\label{39}
\Phi_F(\br) = \dlt(\br) \sum_f 4\pi \; \frac{a_f}{m} \hat P_f \;   ,
\ee
where $a_f$ is the scattering length of a pair of atoms with the angular momentum of 
the pair $f$ and $\hat{P}_f$ is a projection operator onto a state with an even angular 
momentum $f$.

The last term in (\ref{32}) describes dipolar interactions,
\be
\label{40}
  \hat H_D = \frac{\mu_F}{2} \sum_{klmn} \int \psi_k^\dgr(\br) \;  \psi_l^\dgr(\br') 
D_{klmn}(\br-\br') \;  \psi_m(\br')  \psi_n^\dgr(\br) \; d\br d\br' 
\;    
\ee
through the regularized screened dipolar tensor
$$
D_{klmn}(\br) = \Theta(r-b_F) D_{klmn}^0(\br) \exp(-\varkappa_F \br ) \; , 
$$
\be
\label{41}
D_{klmn}^0(\br) =\frac{1}{r^3} \left[\; ({\bf F}_{kn} \cdot {\bf F}_{lm} ) -
3 ({\bf F}_{kn} \cdot \bn ) ({\bf F}_{lm} \cdot \bn ) \; \right] \;  ,
\ee
in which
$$
 r = |\; \br \; | \; , \qquad  \bn = \frac{\br}{r} \; .
$$
The potential regularizing and screening take into account the finite sizes of atoms 
or molecules and the matter around them \cite{Yukalov_40}.

\section{Insulating optical lattices}

Trapped atoms can be loaded into deep optical latices, formed by laser beams, where 
the atoms become well localized and do not jump between lattice sites 
\cite{Morsch_67,Moseley_68,Yukalov_69,Krutitsky_70}. Then the lattice is called 
insulating. For a periodic lattice, the field operators can be expanded over Wannier 
functions,
\be
\label{42}
\psi_m(\br) = \sum_j c_{jm} w(\br-\ba_j) \;   ,
\ee
keeping in mind that at low temperature, the single-band approximation is valid. 
For an insulating lattice, the Wannier functions can be chosen to be well localized 
\cite{Marzari_71}, so that the localization conditions be valid:
$$
\int w^*(\br-\ba_i) \; f(\br) \; w(\br-\ba_j) \; d\br \simeq \dlt_{ij} f(\ba_j) \; ,
$$
$$
\int |\; w(\br-\ba_i)\; |^2 \; f(\br) \; |\; w(\br-\ba_j)\; |^2 \; d\br \simeq 
\dlt_{ij} f(\ba_j) \int  |\; w(\br-\ba_i)\; |^4 \; d\br \;  ,
$$
where $f({\bf r})$ is a smooth function of ${\bf r}$.

Let us introduce the local spin operator
\be
\label{43}
 \bS_j = \sum_{mn} c_{jm}^\dgr \; {\bf F}_{mn} \; c_{jn}  
\ee 
satisfying the standard spin algebra for any statistics of $c_{jm}$, whether Bose or Fermi.
The local density of atoms is defined as
\be
\label{44}
\hat n_j = \sum_m c_{jm}^\dgr \; c_{jn} \; .
\ee

In this notation, the single-atom term (\ref{33}) reads as
\be
\label{45}
\hat H_0 = \sum_j E_j \hat n_j \;   ,
\ee
where 
\be
\label{46}
E_j \equiv \int  w^*(\br-\ba_i) \; \left[\; -\; 
\frac{\nabla^2}{2m} + U(\br) \; \right] \; w(\br-\ba_j) \; d\br \;   .
\ee

The term (\ref{38}) of local atomic interactions, for $F=1$, becomes
\be
\label{47}
 \hat H_F = \sum_j \left[\; \frac{b_0}{2} \; \hat n_j (\hat n_j + 1) +
\frac{b_2}{2} \; \left( \bS_j^2 - 2\hat n_j \right) \; \right] \;  ,
\ee
with the parameters
$$
b_0 = c_0 \int |\; w(\br)\; |^4 \; d\br \; , \qquad
b_2 = c_2 \int |\; w(\br)\; |^4 \; d\br \; , 
$$
\be
\label{48}
c_0 = \frac{4\pi}{3m}\; (a_0 + 2a_2) \; , \qquad 
c_2 = \frac{4\pi}{3m}\; (a_2 - a_0) \;   ,
\ee
in which $a_0$ and $a_2$ are the scattering lengths for collisions of atomic pairs.

The linear Zeeman-effect Hamiltonian (\ref{34}) reduces to
\be
\label{49}
 \hat H_{LZ} = - \mu_F \sum_j \bB_j \cdot \bS_j \;  .
\ee
And the quadratic Zeeman-effect Hamiltonian (\ref{35}) takes the form
\be
\label{50}
\hat H_{QZ} = \sum_j 
\left[\; Q_Z (\bB_j \cdot \bS_j)^2 + q_Z (S_j^z)^2\; \right] \;   .
\ee
Here the local magnetic field is
\be
\label{51}
 \bB_j \equiv \bB(\ba_j) = (B_0 + \Dlt B) \bfe_z + H \bfe_x \;  .
\ee

The dipolar Hamiltonian (\ref{40}) acquires the form
\be
\label{52}
 \hat H_D = \frac{1}{2} 
\sum_{i\neq j} \; \sum_{\al\bt} D_{ij}^{\al\bt} S_i^\al S_j^\bt \;  ,
\ee
with the dipolar tensor
\be
\label{53}
 D_{ij}^{\al\bt}  = \frac{\mu_F^2}{r_{ij}^3} \; 
(\dlt_{\al\bt} - 3 n_{ij}^\al n_{ij}^\bt) \; \exp(-\varkappa r_{ij} ) \;  .
\ee

\section{Spin equations of motion}
 
The spin equations of motion are obtained from the Heisenberg equations
\be
\label{54}
i\; \frac{d S_j^\al}{dt} = [\; S_j^\al , \; \hat H \; ] \;   .
\ee
It is straightforward to notice that 
\be
\label{55}
 [\; S_j^\al , \; \hat H_0 \; ] = 0 \; , \qquad
[\; S_j^\al , \; \hat H_F \; ]  = 0 \; ,
\ee
which is valid for any rotationally symmetric Hamiltonian $\hat{H}_F$ with 
arbitrary $F$. Therefore the spin motion is governed only by the spin part 
of the Hamiltonian, 
\be
\label{56}
 i\; \frac{dS_j^\al}{dt} = [\; S_j^\al , \; \hat H_S \; ] \;   ,
\ee
with the spin Hamiltonian
\be 
\label{57}
\hat H_S = \hat H_{LZ} + \hat H_{QZ} + \hat H_D \;   .
\ee
Notice that the quadratic Zeeman Hamiltonian enters these equations, and its presence is
important for governing spin dynamics \cite{Yukalov_39,Yukalov_40,Yukalov_72,Yukalov_73}.

The spin equations of motion can be analyzed in several ways. First, it is possible to
resort to quasiclassical approximation, when the spin operators $S_j^\alpha$ are replaced
by their averages $\langle S_j^\alpha \rangle$ and the resulting equations are solved 
numerically. This description is applicable at the coherent stage of spin motion, but it 
is not suitable for quantum stages of motion, especially at the initial stage of movement,
where spin waves, triggering the spin motion, are necessary to initiate it.
 
The other method is based on the scale separation approach \cite{Yukalov_45}. Then we write 
down the equations for the averages describing the transverse spin 
\be
\label{58}
u = \frac{1}{SN} \sum_j \lgl \; S_j^- \; \rgl \;   ,
\ee
coherence intensity
\be
\label{59}
 w = \frac{1}{(SN)^2} \sum_{i\neq j} \lgl \; S_i^+ S_j^- \; \rgl \;  ,
\ee
and longitudinal spin polarization
\be
\label{60}
s = \frac{1}{SN} \sum_j \lgl \; S_j^z \; \rgl \;  ,
\ee
where $S_j^{\pm}$ are the ladder spin operators. The expressions describing local dipolar 
fluctuations, responsible for the creation of spin waves, are treated as stochastic 
variables \cite{Yukalov_6,Yukalov_39}.

Keeping in mind the setup, discussed in Sec. 2, the spin equations of motion are 
complemented by the equation (\ref{3}) for the feedback magnetic field $H$ or for its
dimensionless form
\be
\label{61}
 h = - \; \frac{\mu_F H}{\gm_0} \qquad 
( \gm_0 \equiv \pi \eta_{res} \gm_2 ) \;  ,
\ee
with $\gamma_2 = \rho \mu_F^2 S$ and $\rho$ being spin density.

To realize good resonance, the attenuations are to be small as compared to the Zeeman
frequency
\be
\label{62}
\om_0 = - \mu_F B_0 > 0 \;   .
\ee
In that way, there are several small parameters in the system of equations: 
\be
\label{63}
\frac{\gm}{\om_0} \ll 1 \; , \qquad
\frac{\gm_0}{\om_0} \ll 1 \; , \qquad 
\frac{\gm_2}{\om_0} \ll 1 \; , \qquad 
\frac{\gm_3}{\om_0} \ll 1 \; , 
\ee
where the spin-wave attenuation $\gamma_3$ is  
\be
\label{64}
\gm_3 \simeq \frac{\gm_2^2}{\om_0} \;    .
\ee
Spin fluctuations are proportional to $\gamma_2$, hence are also small.

In accordance to the small parameters (\ref{63}), the variables $u$ and $h$ are 
classified as fast, while $w$ and $s$, as slow. Following the scale separation approach, 
that is a variant of the Krylov-Bogolubov averaging technique \cite{Bogolubov_74}, we 
solve the equations for fast variables $u$ and $h$, with slow variables fixed. Then 
we substitute the found fast variables $u$ and $h$ into the equations for the slow 
variables $w$ and $s$ and average the right-hand sides of these equations for the
slow variables over time and random fluctuations, with fixed slow variables $w$ and $s$. 
As a result of this procedure, we obtain the guiding-center equations
$$
\frac{dw}{dt} = - 2 w + 2\al w s + 2 \; \frac{\gm_3}{\gm_2} \; s^2 +
2q \al \bt w^2 s + 2 \; \frac{\om_0}{\gm_2} \; q\al w s^2 \; ,
$$
\be
\label{65}
 \frac{ds}{dt} = - \al w - \frac{\gm_3}{\gm_2} \; s -
q \al \bt w^2 - \; \frac{\om_0}{\gm_2} \; q\al w s \;  ,
\ee 
where time is measured in units of $\gamma_2$. Here $\alpha$ and $\beta$ are coupling 
functions 
$$
\al = \frac{g\gm^2}{\gm^2+\Dlt_S^2} \; ( 1 + b + As) \; \left\{ 1 - 
[\; \cos(\Dlt_S t) - \dlt_S \sin(\Dlt_S t) \; ] \; e^{-\gm t} \right\} \;   , 
$$
\be
\label{66}
\bt = -\; \frac{g\gm^2}{\gm^2+\Dlt_S^2} \; ( 1 + b + As) \; \left\{ \dlt_S - 
[\; \sin(\Dlt_S t) + \dlt_S \cos(\Dlt_S t) \; ] \; e^{-\gm t} \right\} \;   ,   
\ee
with the coupling parameter
\be
\label{67}
 g = \frac{\gm_0\om_0}{\gm\gm_2}  
\ee
and the notation
\be
\label{68}
\Dlt_S = \om - \om_0 |\; 1 + As \; | \; , \qquad
\dlt_S = \frac{\Dlt_S}{\gm} \; {\rm sgn} \; \om_S \;   .
\ee

The quadratic Zeeman-effect parameter
\be
\label{69}
A = \frac{\om_0}{\gm_2} \; q + \frac{\gm_2}{\om_0}\; p
\ee
is composed of the static-current quadratic Zeeman-effect parameter
\be
\label{70}
q = S ( 2S - 1) \rho Q_Z
\ee
and of the alternating-current quadratic Zeeman-effect parameter
\be
\label{71}
 p = ( 2S - 1) \; \frac{q_Z}{\gm_2} \;  .
\ee

The effective spin-rotation Zeeman frequency is
\be
\label{72} 
 \om_S = \om_0 ( 1 + b + As ) \;  ,
\ee
where the regulated field $b=b(t)$ and the alternating-current quadratic Zeeman-effect
parameter $p=p(t)$ can be varied with time. It is therefore possible to regulate the
detuning
\be
\label{73}
\frac{\om_S - \om_0}{\om_0} = b(t) - A(t) s(t)
\ee
so that either to make it large or negligibly small, respectively, either to freeze the 
spin or, by inducing the resonance, to organize a fast spin reversal, as is described
in Sec. 2. The required resonance condition is 
\be
\label{74}
  b(t) - A(t) s(t) = 0 \;  .
\ee

To estimate the typical order of the parameters, let us take the values for the atoms of 
$^7$Li, $^{21}$Na, $^{41}$K, and $^{87}$Rb. The spin density is 
$\rho\sim 10^{15}$ cm$^{-3}$, the attenuation rates are $\gamma_2\sim 10^2$ s$^{-1}$,
$\gamma\sim 10 \gamma_2 \sim 10^3$ s$^{-1}$, for a magnetic field $B_0 \sim 10^4$ G,
we have $\omega_0 \sim 10^{11}$ s$^{-1}$, $g \sim \omega_0/\gamma \sim 10^8$. 

The static-current parameter $Q_Z \sim 10^{-24}$ cm$^3$, and $q \sim \rho Q_Z \sim 10^{-9}$. 
Also, $\omega_0/\gamma_2 \sim 10^9$, hence the contribution of the static-current effect 
into $A$ is $q \omega_0/\gamma_2 \sim 1$. 

In the case of the alternating current effect, $q_Z \sim 10^5$ s$^{-1}$, then
$p \sim q_Z/\gamma_2 \sim 10^3$. Hence the contribution of the alternating-current effect
into $A$ is $p \gamma_2/\omega_0 \sim q_Z \omega_0 \sim 10^{-6}$. Thus, the parameter
$A$ is mainly due to the static-current quadratic Zeeman effect.

\section{Conclusion}

The methods of regulating spin dynamics in different magnetic materials are studied. These 
materials include magnetic nanomolecules, magnetic nanoclusters, magnetic graphene, polarized
nanomolecules, quantum dots, spinor atoms, and hybrid magnetic structures. The basic setup
supposes the use of an electric circuit forming a feedback field acting on the magnetic sample
and efficiently accelerating spin reversal. The ultrafast spin reversal can be realized in
around $10^{-11}$ s. The idea of employing a feedback field, in combination with the suggested
method of triggering resonance, method of dynamic resonance tuning, and the use of the quadratic
Zeeman effect can be applied in spintronics, e.g. for creating memory devices. Since there are 
many similarities between materials with magnetic moments and those possessing electric moments,
the developed ideas and methods can be used for achieving fast reversal of electric dipoles.
For instance, this can be employed for fast reversal of polarization in ferroelectrics coupled
with electric cavity resonators \cite{Yukalov_75,Yukalov_76}.

\newpage

\begin{figure*}[ht]
\centerline{
\includegraphics[width=10cm]{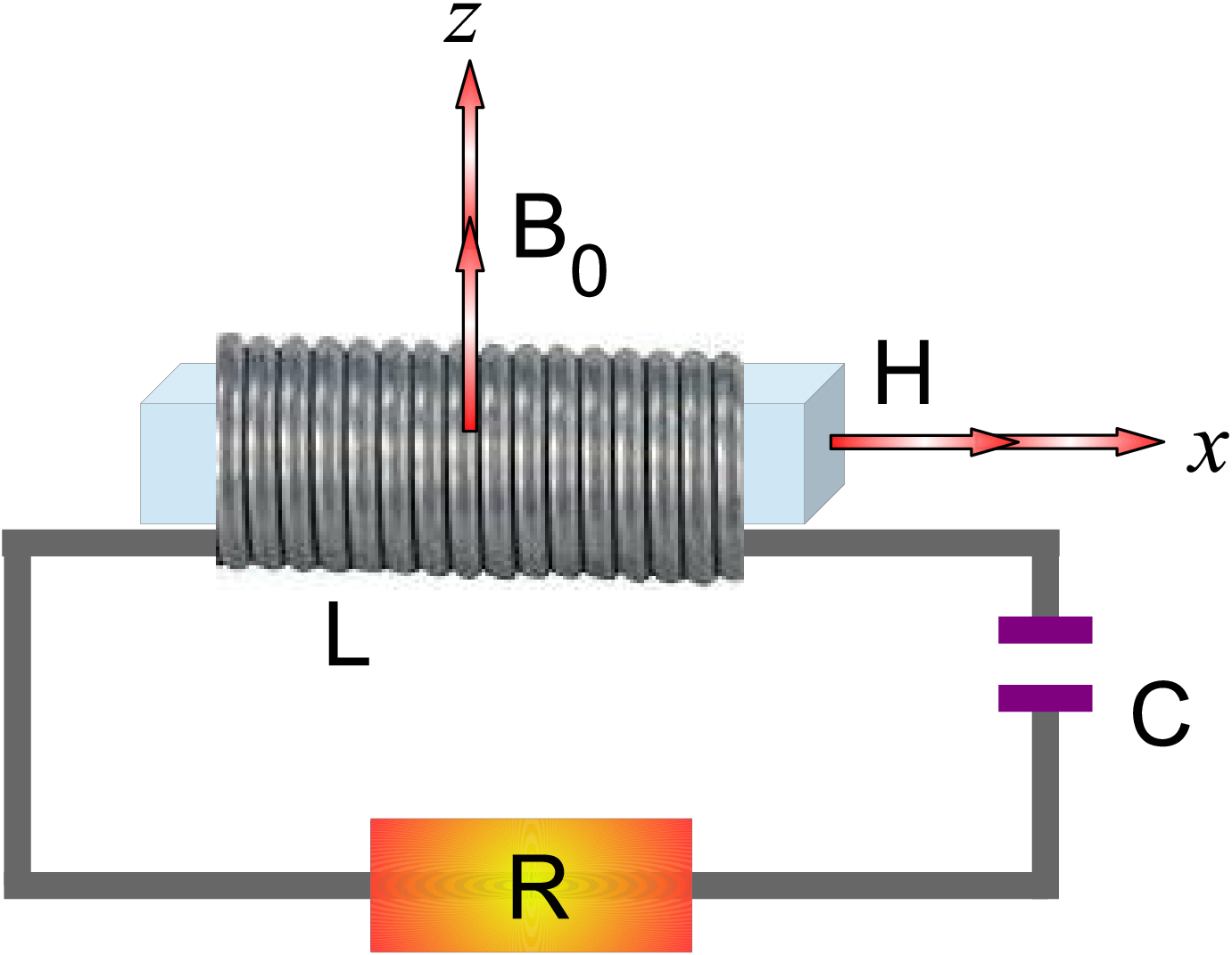} }
\caption{Scheme of suggested setup, as is explained in the
text.}
\label{fig:Fig.1}
\end{figure*}

\begin{figure*}[ht]
\centerline{
\includegraphics[width=10cm]{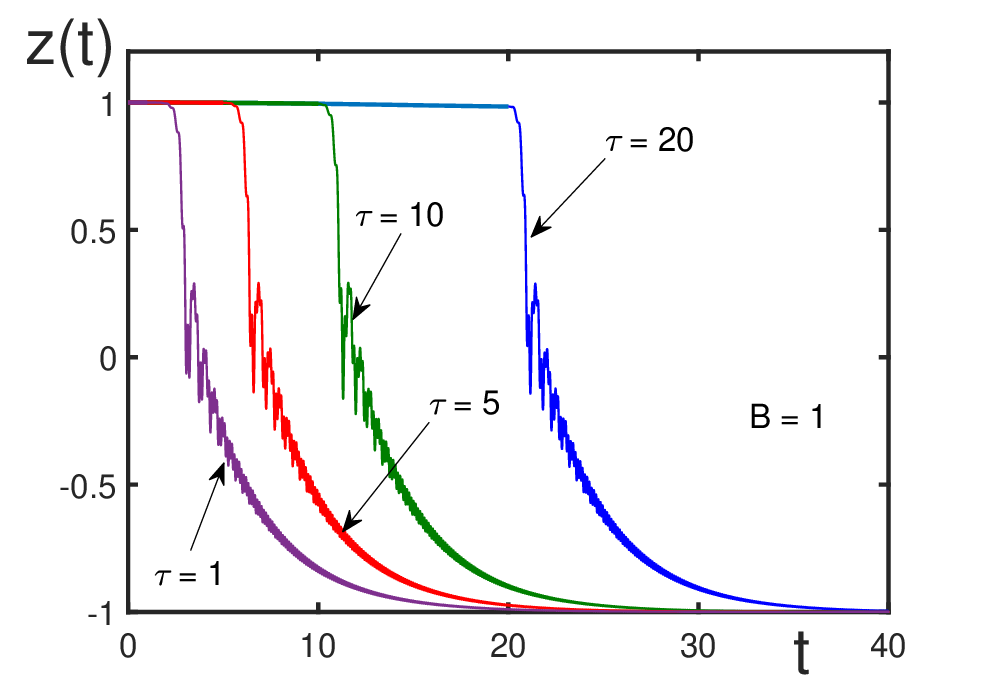} }
\caption{Spin polarization of a nanomagnet as a function of time for the parameters 
$\om=\om_0=10$, $\om_E=\om_1=0.01$, $\gm=1$, with the fixed triggering resonance 
condition $b(t)=A z_0=1$ and varying delay time $\tau$}
\label{fig:Fig.2}
\end{figure*}

\begin{figure*}[ht]
\centerline{
\includegraphics[width=10cm]{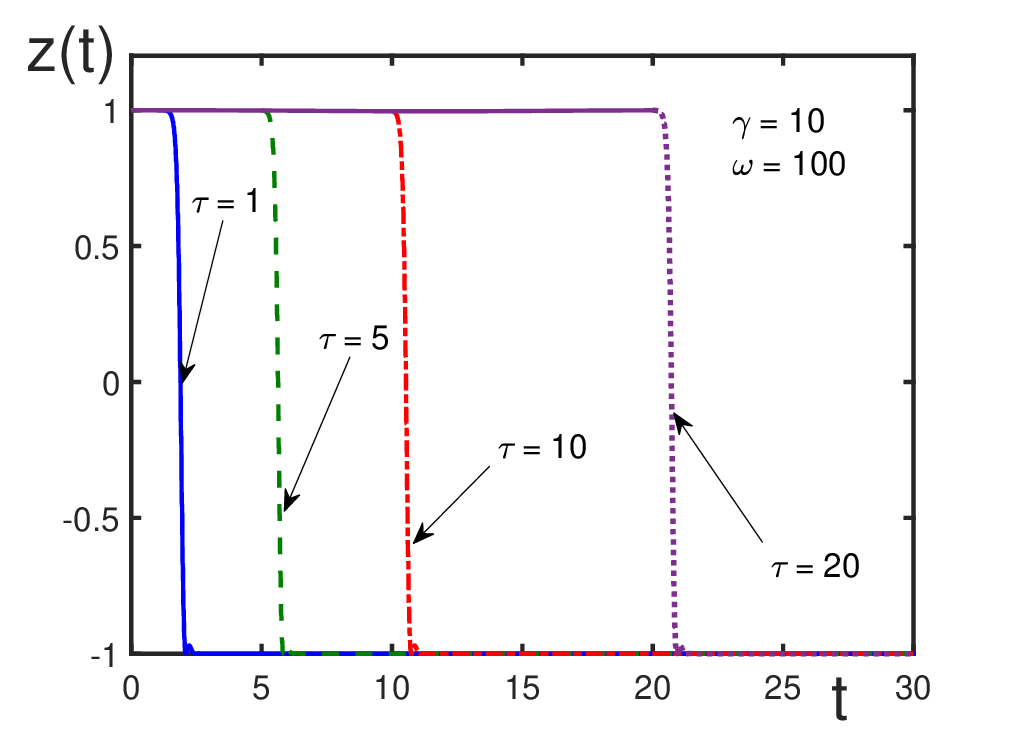} }
\caption{Spin polarization $z(t)$ of a nanocluster or nanomolecule as a function of time 
under dynamic resonance tuning, starting at different delay times: $\tau=1$ (solid line), 
$\tau = 5$ (dashed line), $\tau = 10$ (dash-dotted line), and $\tau = 20$ (dotted line). 
Other parameters are fixed as $A = 1$, $\om_E = \om_1 = 0.01$, $\om = \om_0 = 100$ and 
$\gm = 10$.}
\label{fig:Fig.3}
\end{figure*}

\end{document}